\documentclass[3p,times,twocolumn]{elsarticle}

\usepackage{ecrc}

\volume{00}

\firstpage{1}

\journalname{Nuclear Physics B Proceedings Supplement}

\runauth{P. Hohler and R. Rapp}

\jid{nppp}

\jnltitlelogo{Nuclear Physics B Proceedings Supplement}

\usepackage{amssymb}

\usepackage[figuresright]{rotating}

\def\eg{{\it e.g.}}

\newcommand{\beq}{\begin{equation}}
\newcommand{\eeq}{\end{equation}}
\newcommand{\bea}{\begin{eqnarray}}
\newcommand{\eea}{\end{eqnarray}}

\begin{document}

\begin{frontmatter}



\dochead{}

\title{Dileptons and Chiral Symmetry Restoration}


\author{P. M. Hohler and R. Rapp}
\address{Department of Physics and Astronomy and Cyclotron Institute, Texas A{\&}M
University, College Station, TX 77843-3366, USA}

\begin{abstract}
We report on recent work relating the medium effects observed in dilepton spectra
in heavy-ion collisions to potential signals of chiral symmetry restoration. The key
connection remains the approach to spectral function degeneracy between the vector-isovector
channel with its chiral partner, the axialvector-isovector channel.
Several approaches are discussed to elaborate this connection, namely QCD and Weinberg
sum rules with input for chiral order parameters from lattice QCD, and chiral hadronic
theory to directly evaluate the medium effects of the axialvector channel and the
pertinent pion decay constant as function of temperature. A pattern
emerges where the chiral mass splitting between $\rho$ and $a_1$ burns off and
is accompanied by a strong broadening of the spectral distributions.
\end{abstract}


\end{frontmatter}


\section{Introduction}
\label{sec:intro}
The search for signals of chiral symmetry restoration (CSR) in hot/dense QCD
matter remains one of the key objectives, and challenges, in relativistic
heavy-ion physics. Since the spontaneous breaking of chiral symmetry (SBCS)
in the QCD vacuum is an inherently soft phenomenon, occurring at
scales of around 1~\,GeV and below, low-mass diletpon spectra are arguably
one of the most promising observables to detect changes in the chiral
properties of the finite-temperature/-density ground state. The low-mass
dilepton emissivity, which governs thermal radiation from a hot fireball,
is directly proportional to the hadronic vector ($J^P$=$1^-$) spectral function
of the system. In vacuum, its low-mass spectral strength is concentrated in the
light vector mesons $\rho$, $\omega$ and $\phi$, representing massive hadronic
degrees of freedom as manifestations of the nontrivial condensate structure
of the QCD vacuum. At the quark level, SBCS is associated with the formation
of the quark condensate giving rise to the mass of constituent quarks,
$m_q^*$$\simeq$$- G \langle0|\bar qq|0 \rangle$$\simeq$\,0.3-0.4\,GeV, which in
turn form the building blocks of the hadrons. At the hadronic level,
SBCS emerges as the splitting of chiral multiplets, such as $\pi$-$\sigma$,
$N$-$N^*(1535)$ or $\rho$-$a_1$. Thus, to extract signatures of (the approach
to) CSR from dilepton spectra, which are dominated by the $\rho$ channel, one
is led into investigations of medium effects in the $a_1$ channel.

\begin{figure*}[!t]
\hspace{-0.1cm}
\begin{minipage}{0.22\linewidth}
\vspace{0.1cm}
\includegraphics[width=4.1cm]{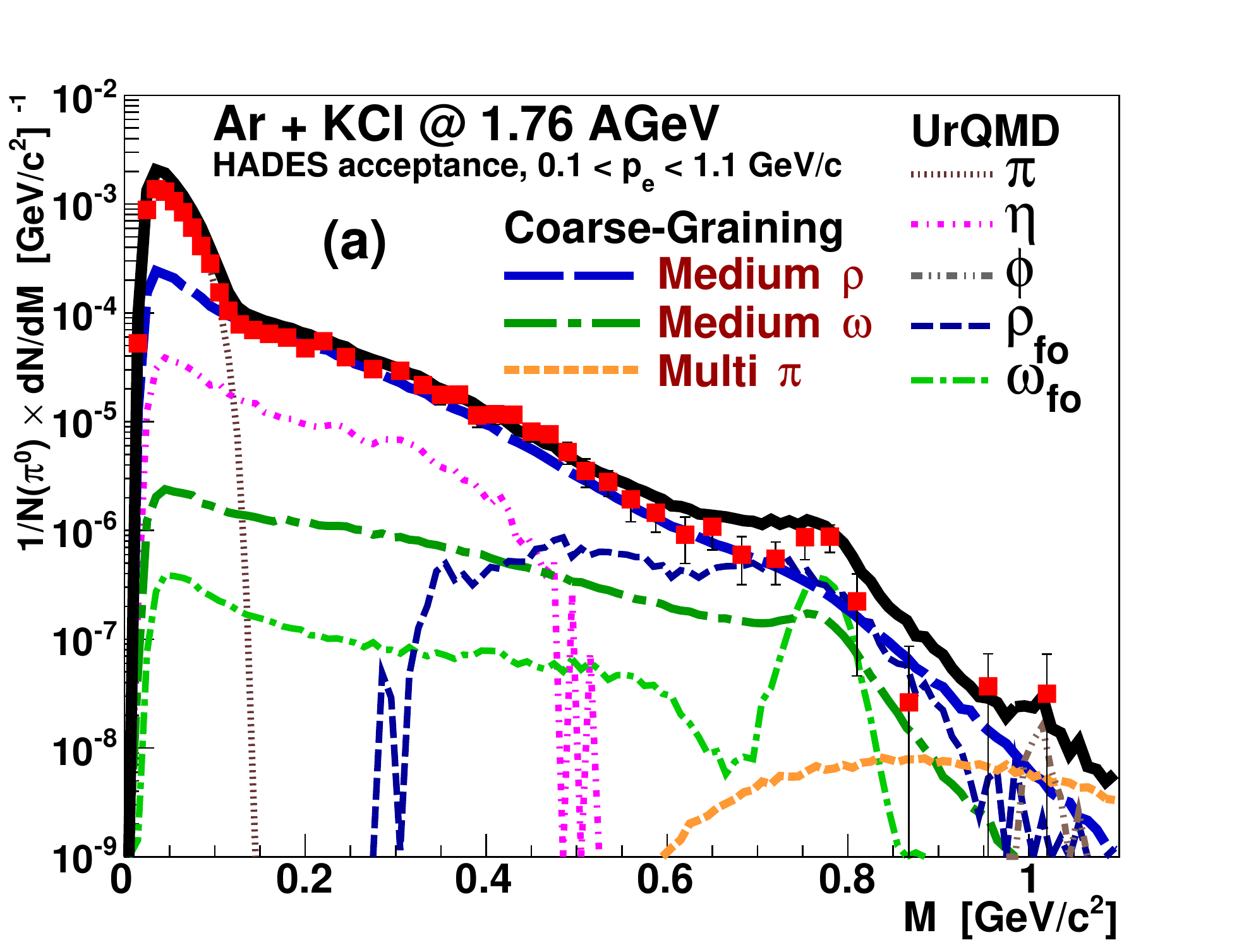}
\end{minipage}
\hspace{-0.15cm}
\begin{minipage}{0.22\linewidth}
\vspace{0.4cm}
\includegraphics[width=4.6cm]{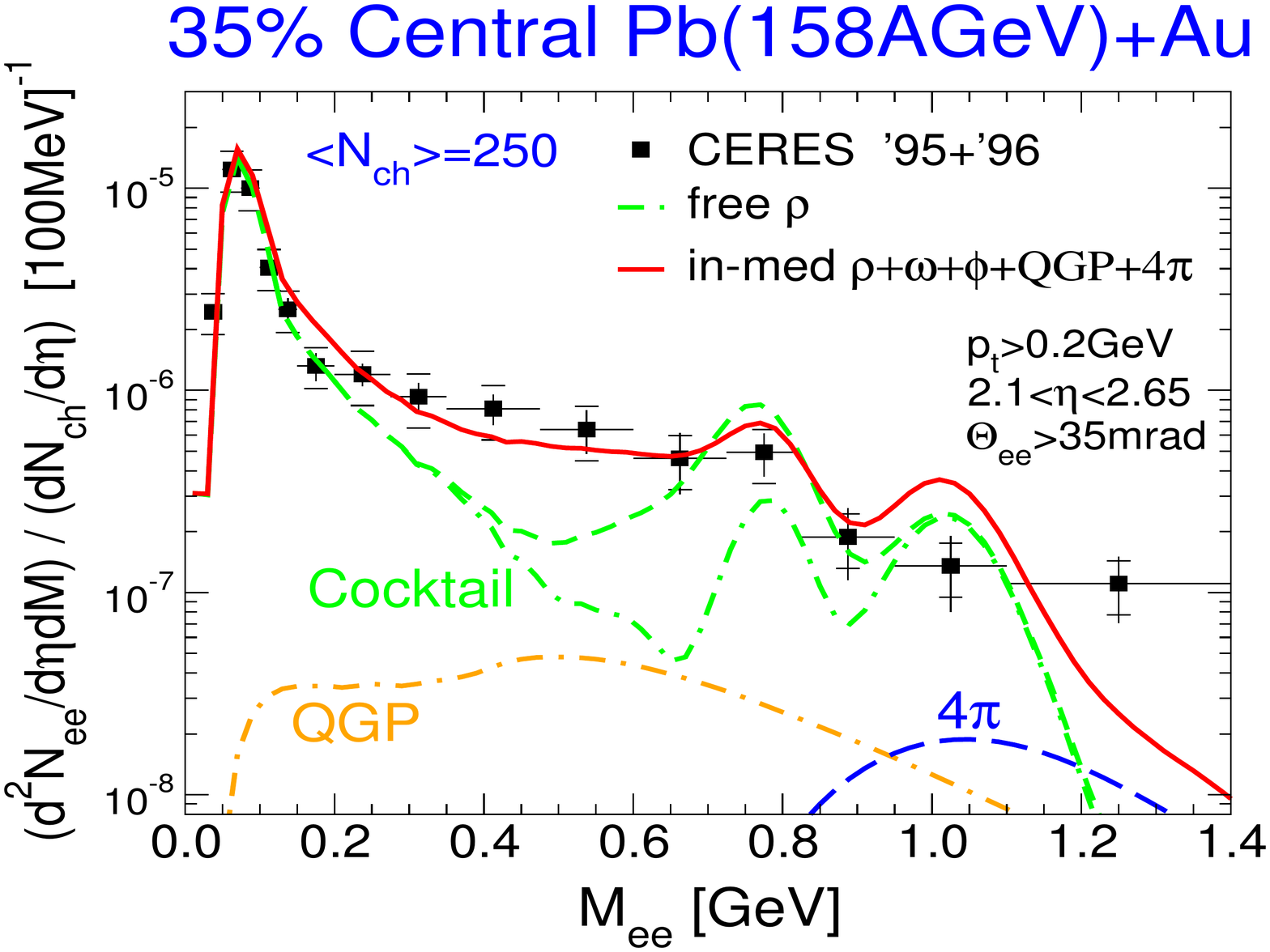}
\end{minipage}
\hspace{0.7cm}
\begin{minipage}{0.22\linewidth}
\vspace{0.6cm}
\includegraphics[width=3.9cm,height=3.1cm]{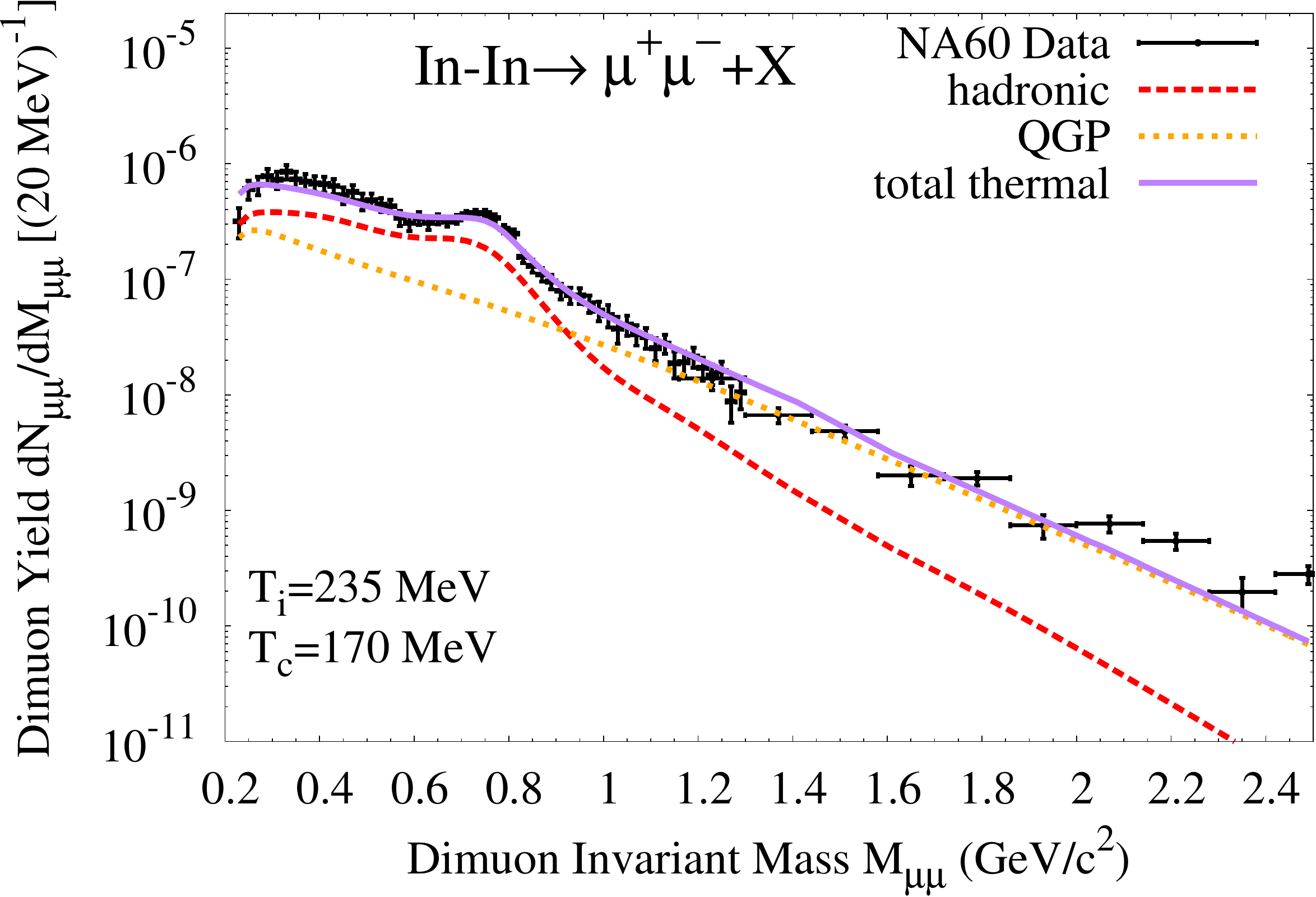}
\end{minipage}
\hspace{0.3cm}
\begin{minipage}{0.22\linewidth}
\includegraphics[width=4.2cm]{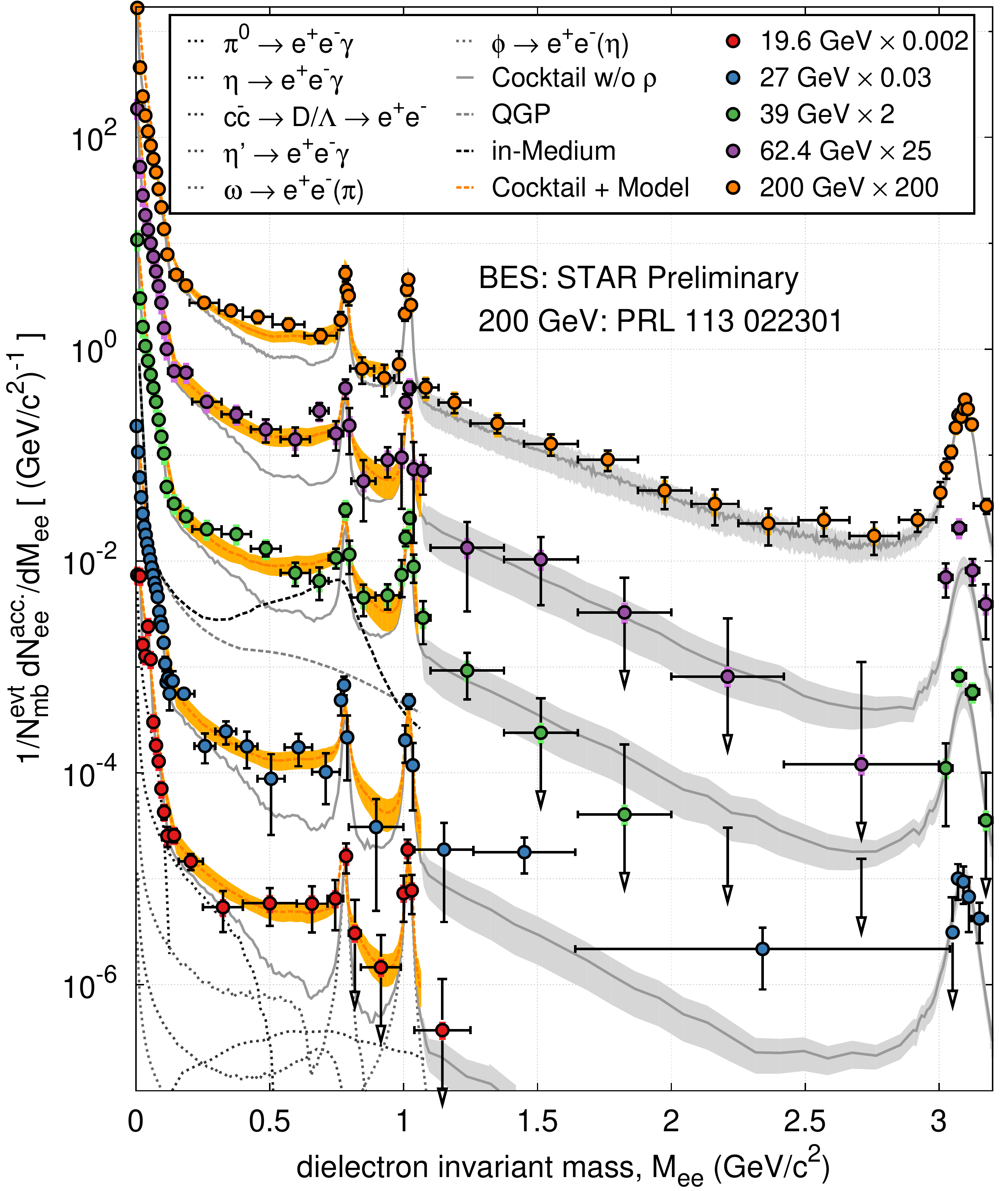}
\end{minipage}
\caption{Dilepton spectra in heavy-ion collisions from HADES at SIS~\cite{Agakishiev:2011vf}
(left), CERES~\cite{Agakichiev:2005ai} and NA60~\cite{Specht:2010xu} at SPS (middle panels)
and STAR~\cite{Huck:2014mfa} at RHIC (right panel), compared to theoretical calculations
with QGP and hadronic emission using the same model for the in-medium $\rho$ spectral
function~\cite{Endres:2015fna,vanHees:2007th,Rapp:2014hha,Rapp:2013nxa}.}
\label{fig:data}
\end{figure*}
The case for investigations of hadronic mechanisms for chiral restoration
in the context of dileptons is further motivated by the following
considerations. From the phenomenological side, it is rather well established
by now that the low-mass enhancement observed in all
experiments, ranging from SIS ($\sqrt{s}$=2.25\,GeV)~\cite{Agakishiev:2011vf}
via SPS ($\sqrt{s}$=8.8,17.3\,AGeV)~\cite{Adamova:2002kf,Agakichiev:2005ai,Specht:2010xu}
via the RHIC beam energy scan to its maximal energy
($\sqrt{s}$=19.6,27,39,62,200\,AGeV)~\cite{Huck:2014mfa,Adamczyk:2015lme},
can be understood via thermal radiation of predominantly hadronic
origin with a strongly broadened $\rho$ spectral function, cf.~Fig.~\ref{fig:data}.
From the theoretical side, one can estimate the energy density corresponding
to hot matter at the pseudo-critical temperature for chiral restoration,
$T_{\rm pc}$$\simeq$155\,MeV (at vanishing baryon chemical potential, $\mu_B$=0).
One finds $\epsilon_{\rm pc}$$\simeq$0.3\,GeV which is comparable to that of cold
nuclear matter at twice saturation density. Of course, one should keep in mind
that the chiral condensate at $T_{\rm pc}$ is still at about half of its vacuum
value, which, again is in the vicinity of recent estimates in cold nuclear matter at
twice nuclear saturation density~\cite{Holt:2013fwa}.

In the following, we will present two frameworks of assessing the in-medium changes
of the chiral properties of the QCD medium in terms of the vector-axialvector spectral
functions. In the first one (Sec.~\ref{sec:srs}), we utilize chiral and QCD sum rules
to search for solutions for the axialvector spectral function given previously
calculated in-medium $\rho$ spectral functions and chiral order parameters as
available from lattice-QCD as input. In the second one (Sec.~\ref{sec:mym}), we compute
vector and axialvector spectral functions in a hot pion gas from a chiral effective theory
and extract the in-medium dependence of the pion decay constant and scalar condensate.
We conclude in Sec.~\ref{sec:concl}.

\section{QCD and Weinberg Sum Rules}
\label{sec:srs}
The QCD sum rule (QCDSR) technique has been widely used to analyze hadronic properties
in both vacuum and medium, by relating the pertinent spectral function to an
expansion in Euclidean momentum transfer where coefficients are given by
vacuum condensates~\cite{Shifman:1978bx}. For the vector 
channel in vacuum, after a Borel transform
to improve convergence, it takes the form
\begin{eqnarray}
\frac{1}{M^2}\!\int_0^\infty \!ds \frac{\rho_{V}(s)}{s} e^{-s/M^2}
  =  \frac{1}{8\pi^2} \left(1+\frac{\alpha_s}{\pi}\right)
+\frac{m_q \langle\bar{q}q\rangle}{M^4}
\nonumber\\
+\frac{1}{24 M^4}\langle\frac{\alpha_s}{\pi} G_{\mu\nu}^2\rangle
- \frac{\pi \alpha_s}{M^6}\frac{56}{81} \langle \mathcal{O}_4^{V} \rangle \ ,
\end{eqnarray}
and similarly for the axialvector channel with sign changes (and slightly different
coefficients) for the chiral odd contributions from the 2- and 4-quark condensates
(additional higher twist contributions arise in a heat bath).
As is well known for the $\rho$ meson~\cite{Hatsuda:1992bv,Leupold:1997dg,Zschocke:2002mn},
the QCDSR cannot uniquely predict the in-medium properties of the spectral function, but
it can provide constraints on model calculations.

Weinberg sum rules (WSRs) relate chiral order parameters to energy moments of
the difference between the vector and axialvector spectral functions~\cite{Weinberg:1967kj},
\begin{equation}
f_n =  \int\limits_0^\infty \frac{ds}{\pi} \ s^n \
 \left[\rho_V(s) - \rho_A(s) \right]  \
\label{wsr}
\end{equation}
with $f_{-1}$=$f_\pi^2$, $f_0$=$f_\pi^2m_\pi^2$=$-2m_q\langle \bar qq\rangle$,
and $f_1$=$-2\pi\alpha_s {O}_4^{\rm SB}$ (the latter is the chirally odd
combination of ${O}_4^{V,A}$). They have been extended to finite temperature in
Ref.~\cite{Kapusta:1993hq}.

\begin{figure*}[!t]
\centering
\includegraphics[width=.8\textwidth]{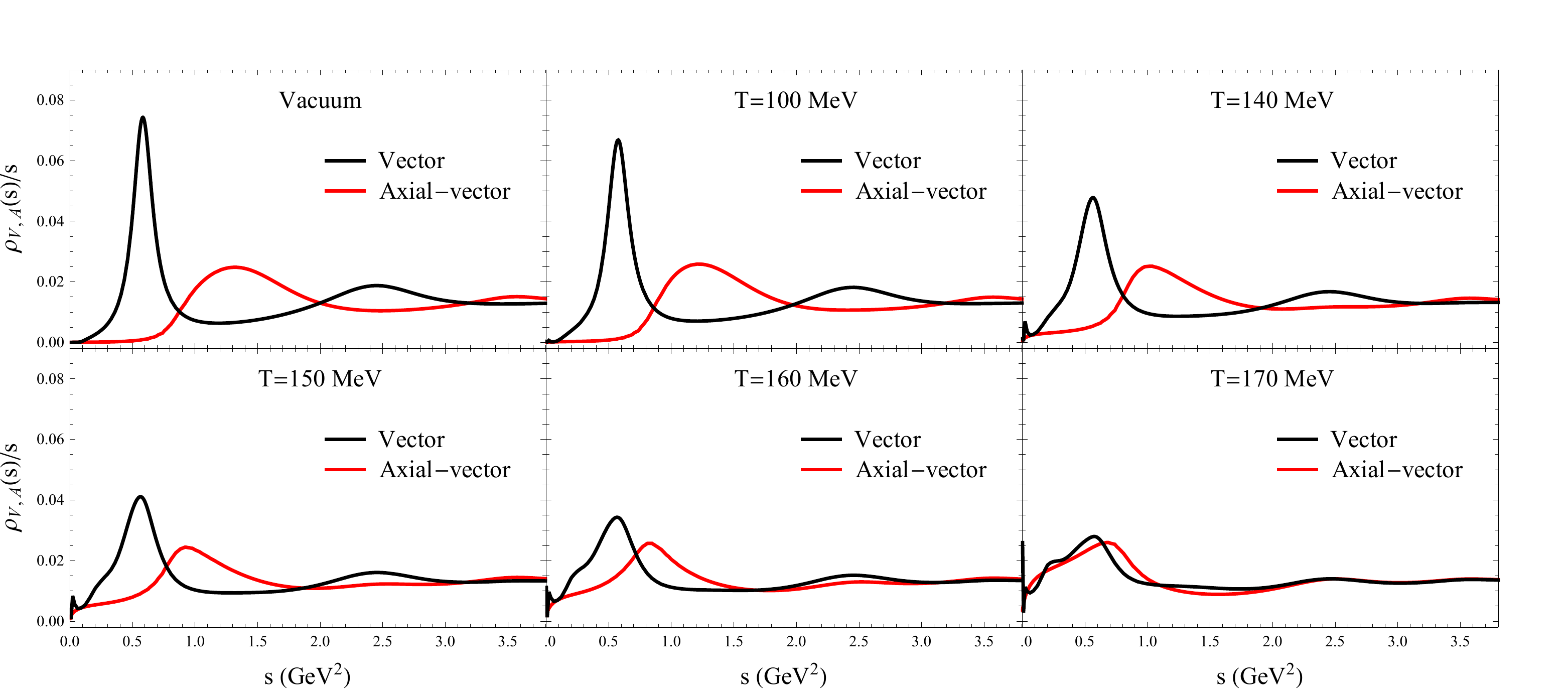}
\caption{Finite-temperature vector (black curve) and axialvector (red curve) spectral functions
from sum rule analysis~\cite{Hohler:2013eba}.}
\label{fig:srs}
\end{figure*}

In Ref.~\cite{Hohler:2013eba}, our idea was to combine the constraining power of QCDSRs
and WSRs to test whether the in-medium $\rho$ spectral function that underlies the
description of dilepton spectra in Fig.~\ref{fig:data} is compatible with chiral
restoration (see also Ref.~\cite{Ayala:2014rka}). As additional input to the sum 
rules the temperature dependence for
$f_\pi(T)$ and $\langle\bar qq\rangle(T)$ was taken from lattice QCD data for the
latter, and inferred via the Gellmann-Oakes Renner relation for the former. The $T$
dependence of the 4-quark condensates is less straightforward. We employed
current algebra for the contributions from the Goldstone bosons and the standard
factorization assumption (valid in a large-$N_c$ expansion) for all other hadrons,
augmented with a $T^{10}$
correction term to render the ${O}_4$'s vanishing at the same temperature as the
$\langle\bar qq\rangle$ condensate. The question was then whether an in-medium
axialvector spectral function could be found to satisfy the 2 QCDSRs and 3 WSRs.
As a starting point, we employed a previously constructed quantitative description of
the axial-/vector spectral functions in vacuum as measured in hadronic $\tau$ decays,
which, in addition to the $\rho$ and $a_1$, include one excited state in each channel
($\rho'$ and $a_1'$) and a chirally invariant (degenerate) continuum~\cite{Hohler:2012xd}.

First, we tested the the QCDSR for the vector channel. It turned out that this was
satisfied well within the typical 1\% benchmark, including a small correction to
the vector dominance coupling constant being reduced by around 5\%. To facilitate
the search for an in-medium axialvector spectral function, we augmented the vacuum
fit by a 4-parameter ansatz for the in-medium changes of mass, width (2),  and coupling
strength, and treated the $T$ dependence of the excited states via chiral mixing.
We then searched for minima in the quadratic sum of the deviations of the axialvector
QCDSR and the first and second WSR ($n$=-1 and 0 in Eq.~(\ref{wsr}); the 3.~WSR is beset
with larger uncertainty already in vacuum but still satisfied in medium at a comparable
level). A smoothly varying in-medium axialvector spectral function could be found which varies smoothly
with temperature and satisfies the SRs at the same level of accuracy as in vacuum.
The result is compared to the vector spectral function in Fig.~\ref{fig:srs}, showing
their progression toward degeneracy. This study shows that the $\rho$ spectral function
underlying the description of dilepton spectra is compatible with the dropping quark
condensate computed in lattice QCD and suggests that chiral restoration is satisfied
by a combined broadening and burning-off of the $\rho$-$a_1$ mass splitting.

\section{Massive-Yang Mills Chiral Lagrangian}
\label{sec:mym}
The local-gauge procedure was the method of choice to introduce axial-/vector
mesons into chiral pion Lagrangians for many years, via Massive Yang-Mills~\cite{Gomm:1984at}
or Hidden Local Symmetry~\cite{Bando:1987br,Harada:2003jx}
approaches. However, it turned out to be rather challenging to quantitatively
describe the vacuum axial-/vector spectral functions in these approaches, as the
single gauge coupling constant does not seem to generate enough strength in the
axialvector channel. In Ref.~\cite{Hohler:2013ena}, we found that the implementation
of a resummed (broad) $\rho$ spectral function into the vacuum $a_1$ selfenergy,
along with (rather involved) vertex corrections required to maintain chiral symmetry,
can largely overcome this problem and yield fair description of the vacuum data.

\begin{figure*}[!t]
  \centering
        \includegraphics[width=1\textwidth]{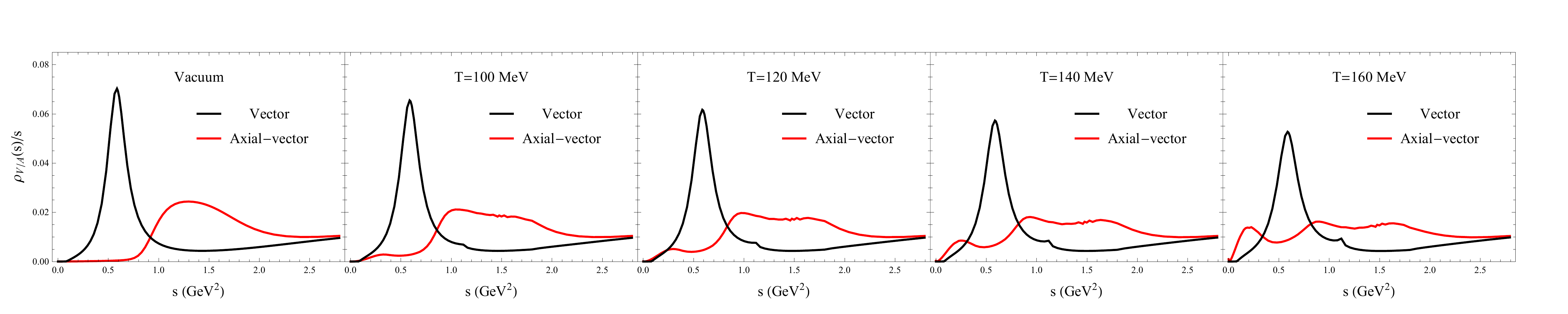}
\caption{Finite-$T$ vector and axialvector spectral functions within
the non-linear realization of Massive Yang-Mills~\cite{Hohler:2015}.}
\label{fig:mym}
\end{figure*}

The next step is to implement this approach at finite temperature, which we have
carried out in recent work for a pion gas~\cite{Hohler:2015}. The vector and
axialvector selfenergies
have been evaluated within the Matsubara formalism accounting for medium effects
through unitarity cuts (Bose enhancement in decay diagrams such as $\rho\to\pi\pi$
and $a_1\to\pi\rho$),  Landau cuts (scattering diagrams with an incoming thermal
pion such as $\pi\rho\to a_1$ or $\pi a_1\to\rho$), vertex corrections and tadpole
diagrams. We have verified that the model-independent low-temperature chiral properties
(\eg, no axial/vector meson mass shifts at order ${\cal O}(T^2)$ in the chiral limit)
are satisfied within our approach.
The resulting temperature progression of the vector and axialvector
spectral functions is shown in Fig.~\ref{fig:mym}. The $\rho$ resonance only
undergoes a moderate broadening of up to $\sim$50\,MeV at $T$=160\,MeV, which
is in line with previous pion gas studies for Bose enhancement and $\pi\rho\to a_1$
resonance excitations. The $a_1$ resonance exhibits stronger modifications,
by developing a larger in-medium width, a reduction in its mass and a low-mass
excitation caused by the $\pi a_1\to\rho$ excitation (``chiral mixing").
These features are qualitatively similar to what we found within the sum rule
analysis discussed in the previous section. They are also rather robust between
employing a linear or non-linear realization of chiral symmetry in the pion
Lagrangian (the former involves an additional $\sigma$ field).

The main benefit of a microscopic chiral approach is that one can utilize
it to calculate chiral order parameters. We have evaluated both the pion
decay constant, $f_\pi(T)$, and, within the linear realization, the scalar
condensate, $\sigma(T)$, and found that both quantities drop by up to
15-20\% at $T$=160\,MeV. This corroborates that the trend toward degeneracy
in the spectral functions is indeed coupled to an approach toward chiral
restoration.

\section{Conclusions}
\label{sec:concl}
We have analyzed the in-medium properties of $\rho$ and $a_1$ in the context
of the approach  toward chiral restoration. First, evaluating QCD and Weinberg
sum rules with a realistic in-medium $\rho$ spectral function and condensates
taken from lattice QCD, we found a solution for the axialvector spectral function
which smoothly degenerates with the $\rho$ via a combined broadening and burning-off of
the chiral mass splitting. We then utilized a Massive-Yang Mills approach for
$\rho$ and $a_1$ mesons with realistic vacuum spectral functions to evaluate
medium effects in a hot pion gas. Also here, a spectral broadening was accompanied
by a reduction in the $a_1$ mass. This trend toward spectral degeneracy is coupled
to a reduction in the pion decay constant evaluated within the same approach.
While baryons are not yet included in this calculation, it is interesting to
note that a recent study of the correlation functions of the nucleon and its
chiral partner, $N^*(1535)$, at finite temperature found the latter's mass to
drop toward, and ultimately degenerate with, the nucleon mass~\cite{Aarts:2015mma}.
This is quite in line with our findings for the vector-axialvector channels,
and suggests a general hadronic mechanism of chiral restoration, where the
mass splitting in chiral multiplets burns off while the ground-state masses
are rather stable with temperature. Indeed, the latter feature is quantitatively
consistent with the observed spectral shape of thermal dilepton radiation in
heavy-ion collisions, recall Fig.~\ref{fig:data}.
\\

{\bf Acknowledgments}.
This work has been supported by the US-NSF under grant No.~PHY-1306359.



\begin{thebibliography}{00}

\bibitem{Borsanyi:2010bp}
  S.~Borsanyi {\it et al.},
  JHEP {\bf 1009} (2010) 073.


\bibitem{Bhattacharya:2014ara}
  T.~Bhattacharya {\it et al.},
  Phys.\ Rev.\ Lett.\ {\bf 113} (2014)  082001.


\bibitem{Agakishiev:2011vf}
  G.~Agakishiev {\it et al.} [HADES Collaboration],
  Phys.\ Rev.\ C {\bf 84}  (2011) 014902.

\bibitem{Adamova:2002kf}
  D.~Adamova {\it et al.} [CERES/NA45 Collaboration],
  Phys.\ Rev.\ Lett.\  {\bf 91} (2003) 042301.

\bibitem{Agakichiev:2005ai}
  G.~Agakichiev {\it et al.} [CERES Collaboration],
  Eur.\ Phys.\ J.\ C {\bf 41} (2005) 475.

\bibitem{Specht:2010xu}
  H.J.Specht [NA60 Collaboration],
  AIP Conf.\,Proc.\,{\bf 1322} (2010) 1.

\bibitem{Huck:2014mfa}
  P.~Huck [STAR Collaboration],
  Nucl.\ Phys.\ A {\bf 931} (2014) 659.

\bibitem{Adamczyk:2015lme}
  L.~Adamczyk {\it et al.} [STAR Collaboration],
  Phys.\ Rev.\ C {\bf 92} (2015) 024912.

\bibitem{Endres:2015fna}
  S.~Endres, H.~van Hees, J.~Weil and M.~Bleicher,
  Phys.\ Rev.\ C {\bf 92} (2015) 1,  014911.

\bibitem{vanHees:2007th}
  H.~van Hees and R.~Rapp,
  Nucl.\ Phys.\ A {\bf 806} (2008) 339.

\bibitem{Rapp:2014hha}
  R.~Rapp and H.~van Hees,
  arXiv:1411.4612 [hep-ph].

\bibitem{Rapp:2013nxa}
  R.~Rapp,
  Adv.\ High Energy Phys.\  {\bf 2013} (2013) 148253.




\bibitem{Holt:2013fwa}
  J.W.~Holt, N.~Kaiser and W.~Weise,
  Prog.\ Part.\ Nucl.\ Phys.\  {\bf 73} (2013) 35.

\bibitem{Shifman:1978bx}
  M.A.~Shifman, A.I.~Vainshtein and V.I.~Zakharov,
  Nucl.\ Phys.\ B {\bf 147} (1979) 385; 448.

 
\bibitem{Hatsuda:1992bv}
  T.~Hatsuda, Y.~Koike and S.-H.~Lee,
  Nucl.\ Phys.\ B {\bf 394} (1993) 221.

\bibitem{Leupold:1997dg}
  S.~Leupold, W.~Peters and U.~Mosel,
  Nucl.\ Phys.\ A {\bf 628} (1998) 311.

\bibitem{Zschocke:2002mn}
  S.~Zschocke, O.P.~Pavlenko and B.~K\"ampfer,
  Eur.\ Phys.\ J.\ A {\bf 15} (2002) 529.
  
 \bibitem{Weinberg:1967kj}
  S.~Weinberg,
  Phys.\ Rev.\ Lett.\  {\bf 18} (1967) 507.  

\bibitem{Kapusta:1993hq}
  J.I.~Kapusta and E.V.~Shuryak,
  Phys.\ Rev.\ D {\bf 49} (1994) 4694.

\bibitem{Hohler:2013eba}
  P.M.~Hohler and R.~Rapp,
  Phys.\ Lett.\ B {\bf 731} (2014) 103.

\bibitem{Ayala:2014rka}
  A.~Ayala, C.A.~Dominguez, M.~Loewe and Y.~Zhang,
  Phys.\ Rev.\ D {\bf 90} (2014) 034012.

\bibitem{Hohler:2012xd}
  P.M.~Hohler and R.~Rapp,
  Nucl.\ Phys.\ A {\bf 892} (2012) 58.

\bibitem{Gomm:1984at}
  H.~Gomm, O.~Kaymakcalan and J.~Schechter,
  Phys.\ Rev.\ D {\bf 30} (1984) 2345.

\bibitem{Bando:1987br}
  M.~Bando, T.~Kugo and K.~Yamawaki,
  Phys.\ Rept.\  {\bf 164}  (1988) 217.

\bibitem{Harada:2003jx}
  M.~Harada and K.~Yamawaki,
  Phys.\ Rept.\  {\bf 381} (2003) 1.

\bibitem{Hohler:2013ena}
  P.M.~Hohler and R.~Rapp,
  Phys.\ Rev.\ D {\bf 89} (2014) 125013.

\bibitem{Hohler:2015}
P.M.~Hohler and R.~Rapp, in preparation.


\bibitem{Aarts:2015mma}
  G.~Aarts, C.~Allton, S.~Hands, B.~Jaeger, C.~Praki and J.~I.~Skullerud,
  Phys.\ Rev.\ D {\bf 92} (2015) 014503.

 \end{thebibliography}



\end{document}